\documentclass[aps,prb,twocolumn,showpacs,superscriptaddress,groupedaddress]{revtex4}
\usepackage{graphicx}
\usepackage{dcolumn}
\usepackage{bm}
\usepackage{amssymb}
\newcommand{\etal}{\textit{et al.}}
\begin{document}
\title{Magnetic structure of GdBiPt: A candidate antiferromagnetic topological insulator}

\author{ R. A. M\"uller} \affiliation{D\'epartement de physique, Universit\'e de Montr\'eal,
Montr\'eal, QC, Canada}\altaffiliation{Regroupement Qu\'eb\'ecois sur les
Mat\'eriaux de Pointe (RQMP)}
\author{N. R. Lee-Hone}
\affiliation{Department of Physics, McGill University, 3600 University St.,
Montr\'eal, QC, Canada}\altaffiliation{Regroupement Qu\'eb\'ecois sur les
Mat\'eriaux de Pointe (RQMP)}
\author{L. Lapointe} \affiliation{D\'epartement de physique, Universit\'e de Montr\'eal,
Montr\'eal, QC, Canada}\altaffiliation{Regroupement Qu\'eb\'ecois sur les
Mat\'eriaux de Pointe (RQMP)}
\author{D. H. Ryan}
\affiliation{Department of Physics, McGill University, 3600 University St.,
Montr\'eal, QC, Canada}\altaffiliation{Regroupement Qu\'eb\'ecois sur les
Mat\'eriaux de Pointe (RQMP)}
\author{T.~Pereg-Barnea}
\affiliation{Department of Physics, McGill University, 3600 University St.,
Montr\'eal, QC, Canada}\altaffiliation{Regroupement Qu\'eb\'ecois sur les
Mat\'eriaux de Pointe (RQMP)}
\author{A. D. Bianchi}\affiliation{D\'epartement de physique,Universit\'e de Montr\'eal,
Montr\'eal, QC, Canada}\altaffiliation{Regroupement Qu\'eb\'ecois sur les
Mat\'eriaux de Pointe (RQMP)}
\author{Y.~Mozharivskyj}\affiliation{Department of Chemistry and Chemical Biology
McMaster University, Hamilton, ON, Canada}
\author{R.~Flacau} \affiliation{Canadian Neutron Beam Centre,
Chalk River Laboratories, ON, Canada}

\date{\today}

\begin{abstract}
A topological insulator is a state of matter which does not break any symmetry
and is characterized by topological invariants, the integer expectation values
of non-local operators.  Antiferromagnetism on the other hand is a broken
symmetry state in which the translation symmetry is reduced and time reversal
symmetry is broken. Can these two phenomena coexist in the same material? A
proposal by Mong {\it et al.}\cite{Mong2010} asserts that the answer is yes.
Moreover, it is theoretically possible that the onset of antiferromagnetism
enables the non-trivial topology since it may create spin-orbit coupling effects
which are absent in the non-magnetic phase.  The current work examines a real
system, half-Heusler GdBiPt, as a candidate for topological antiferromagnetism.
We find that the magnetic moments of the gadolinium atoms form ferromagnetic
sheets which are stacked antiferromagnetically along the body diagonal.  This
magnetic structure may induce spin orbit coupling on band electrons as they hop
perpendicular to the ferromagnetic sheets.  \end{abstract}

\maketitle

The discovery of the quantum Hall effect (QHE) \cite{Klitzing1980} led to a new
way of classifying matter - a phase transition does not have to be bound to
spontaneous symmetry breaking. Two years after von Klitzing's discovery,
Thouless, Kohmoto, Nightingale and den Nijs (TKNN) \cite{Thouless1982}
developed the concept of topological invariants in their description of the same
effect. The TKNN number represents the topology of the system in the form of an
integral of the Bloch wave functions over the Brillouin zone.  This non-local
operation results in an integer number which also corresponds to the number of
dissipation-less edge modes.  The edge modes are guaranteed by the topology and
are also protected by it.  The TKNN number, however, explicitly breaks time
reversal symmetry and is therefore zero in a time reversal invariant system.

In 2005, Kane and Mele~\cite{KaneMele1} proposed a new state of
matter: A topological system which does not break time reversal (TR)
symmetry.  In their example, the topological number is defined modulo
2 and the edge modes gives rise to the quantum spin Hall effect.  Many
exciting developments have been presented since, both experimentally
and theoretically but many questions still remain~\cite{Bernevig2006,
Moore2010, Hasan2010}.  One such question is whether the topological
order can coexist with a broken symmetry state. Moreover, is it
possible for a local order parameter which breaks one or more
symmetries to give rise to topological order?  The answer to
this question, theoretically, is a tentative
`yes'~\cite{Mong2010,Fang2013}.

\begin{figure}
\includegraphics[width=\linewidth]{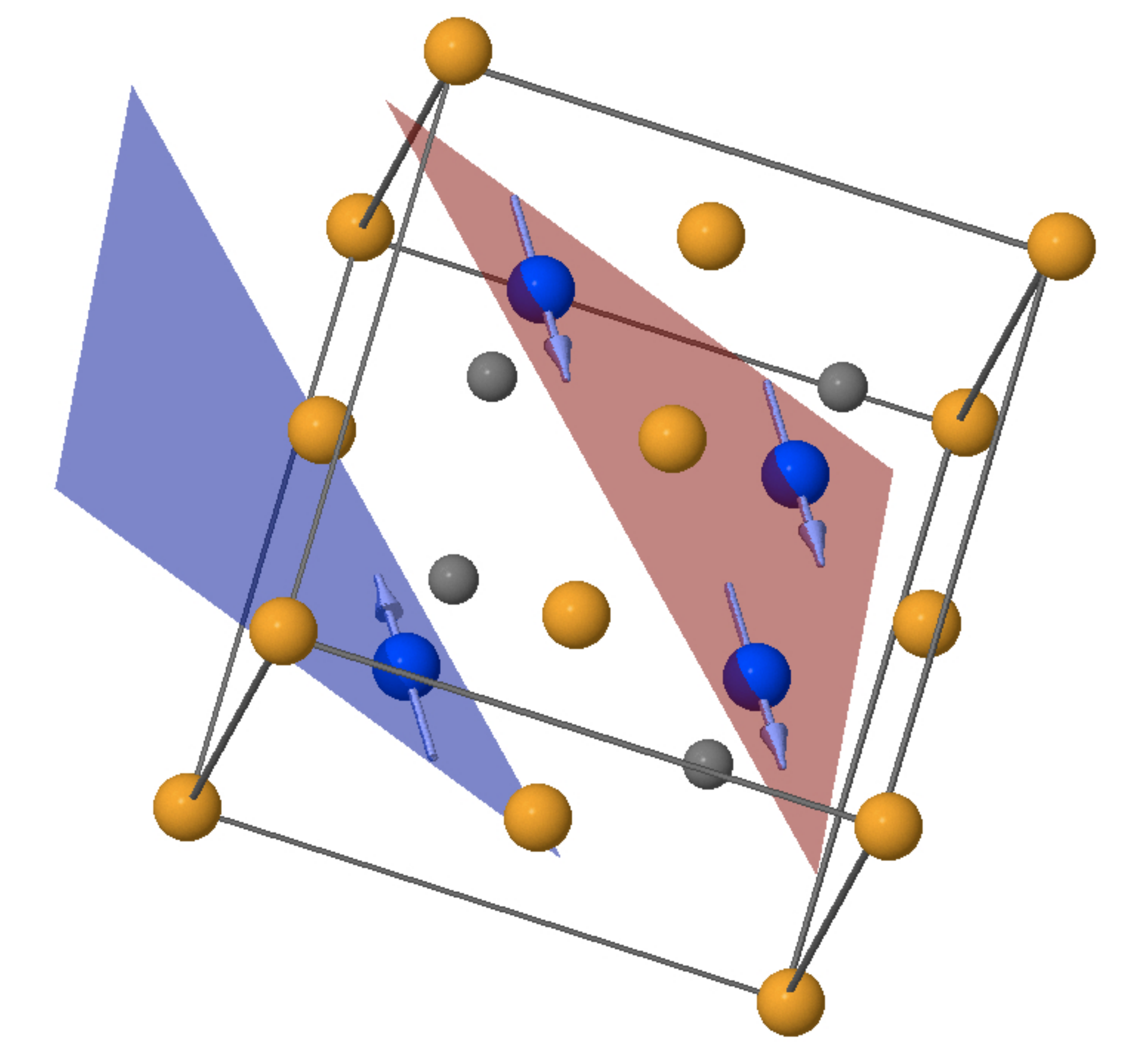}
\caption{The Gd atoms are shown in black (blue), the Bi as gray (gray), and the
Pt as white (yellow). The spins on the Gd atoms are oriented in ferromagnetic
planes which are stacked antiferromagnetically along the magnetic propagation
vector $(\frac{1}{2} \frac{1}{2} \frac{1}{2})$.  \label{fig:sheets}}
\end{figure}

In 2010 Mong {\it et al.}\cite{Mong2010} came forward with the concept of an
antiferromagnetic topological insulator (AFTI). In contrast to an
\emph{ordinary} topological insulator, in an AFTI the presence of magnetic order
breaks TR symmetry $\Theta$
as well as primitive-lattice translational symmetry $T_{1/2}$, yet their product
$S=\Theta~T_{1/2}$ is preserved.  This allows the definition of a topological
invariant which preserves the $S$ symmetry.  In three dimensions the result is a
topological state with antiferromagnetic order.  Depending on whether the
surface breaks the $S$ symmetry or not, metallic surface states may arise within
the band gap and a half-integer quantum Hall effect is expected\cite{Mong2010}.
Moreover, in certain systems, the presence of the topological phase is bound to
the antiferromagnetic phase and so vanishes above the N\'eel temperature. This
makes the AFTI particularly interesting, as the topological state appears only
after the system undergoes a classical phase transition. Therefore, changing the
temperature allows one to turn the topological state on and off resulting in a
quantum phase transition at $T_\mathrm{N}$. Mong {\it et al.}\cite{Mong2010}
propose in their ``model B''
that the spin-orbit interaction may result from the N\'eel order. Their model
contains itinerant electrons and fixed spins.  When the electrons hop between
lattice sites they may do so through intermediate magnetic sites.  For certain
paths of the conduction electrons the magnetic moments serve to create an
Aharonov-Bohm-like flux which in turn acts as Rashba spin-orbit coupling,
responsible for the topological order.  The theoretical model is inspired by
systems like GdBiPt which have been proposed to be topological based on first
principles calculations~\cite{Chadov2010,Lin2010,Xiao2010}.  In order for
the $S$ symmetry to be preserved together with a significant spin orbit coupling
the model requires a specific magnetic structure.  The moments should be aligned
ferromagnetically in layers which are stacked antiferromagnetically.  For the
system to be gapped, the hopping between layers should be larger than the
hopping within the layer.  For the half-Heusler structure, this spin-orbit term
is maximal if the moments are aligned ferromagnetically in the $(111)$ plane and
stacked antiferromagnetically along the $[111]$ space diagonal as
shown in Figure \ref{fig:sheets}  \cite{Mong2010}. The Heusler and the derivative half-Heusler
structures favour half-metallic band structures with just one band crossing at
the Fermi level, while leaving all the other bands well separated and have been
also proposed as candidate materials for \emph{conventional} topological
insulators \cite{Chadov2010,Al-Sawai2010}. The purpose of the current work is to
test whether the desired magnetic structure does indeed occur in GdBiPt. We report on
powder neutron scattering measurements of GdBiPt which shows a magnetic
structure very similar to the one proposed in \cite{Mong2010}, with the magnetic
moments arranged in ferromagnetic sheets, perpendicular to the $[111]$ space
diagonal. This makes GdBiPt a strong candidate for this new state of matter.

GdBiPt crystallizes in the cubic half-Heusler crystal structure with the space
group $F\bar{4}3m$ \cite{Canfield1991}. Members of the $RE$BiPt family show many
interesting properties such as superconductivity, antiferromagnetic order and
super-heavy-fermion behaviour. Band structure calculations and ARPES experiments
on Lu, Nd, and GdBiPt \cite{Liu2011} indicate the presence of metallic surface
states that differ strongly from the band structure in the bulk. However, the
authors found that within their resolution an even number of bands cross the
Fermi level at the surface, making these states sensitive to disorder
unlike in strong topological insulators where an odd number of crossings is
expected, protecting surface states from being backscattered by a non-magnetic
impurity. An X-ray resonant magnetic scattering (XRMS) study on GdBiPt indicated a
doubling of the unit cell along its $[111]$ space diagonal, however the authors
were unable to establish the exact direction of the magnetic moments
\cite{Kreyssig2011}, information that is essential in determining whether
GdBiPt could be an AFTI.

The half-Heusler structure consists of four interpenetrating fcc lattices
shifted by $[\frac{1}{4},\frac{1}{4},\frac{1}{4}]$, three of them occupied by a
different element while the fourth forms an ordered vacancy. We carried out
combined refinement of our X-ray and neutron scattering data, which yields the
lowest $\chi^2$, if the atoms in GdBiPt take the same positions as reported for
YbBiPt  \cite{Robinson1994} and CeBiPt \cite{Wosnitza2006} - platinum  located
on the $[0,0,0]$ site ($4a$), Gd$^{3+}$ on the
$[\frac{1}{4},\frac{1}{4},\frac{1}{4}]$ ($4c$), and Bi on the
$[\frac{3}{4},\frac{3}{4},\frac{3}{4}]$ position ($4d$) (See Table I
of \cite{SOM}). These atomic positions are in agreement
with the ones that have been previously reported by Kreyssig \etal\
\cite{Kreyssig2011}. In addition, we also carried out a single crystal X-ray
diffraction experiment. Due to the non-centrosymmetric nature of the $F\overline{4}3m$
space group, we also tested an inverted structure (racemic twin) with Pt on the
$4a$, Bi on the $4c$ and Gd on the $4d$ site in order to see if such a structure
could account for the observed intensities. In a non-centrosymmetric
structure, anomalous  X-ray  scattering leads to 
different intensities for so-called  \emph{Friedel} pairs, such as
$(hkl)$ and $(\bar{h}\bar{k}\bar{l})$. The refinement confirmed the
original structure, resulting in R1 = 0.0241, where R1 is the
difference between the experimental observations and the ideal
calcluated values, and a Flack parameter, which is the
absolute structure factor,  of -0.13(2) for the current structure in contrast to
R1 = 0.0806 and Flack parameter of 1.2(1) for the inverted structure (please
note that a Flack parameter is 0 for the correct structure and 1 for the
inverted structure).

\begin{figure}
\includegraphics[width=\linewidth]{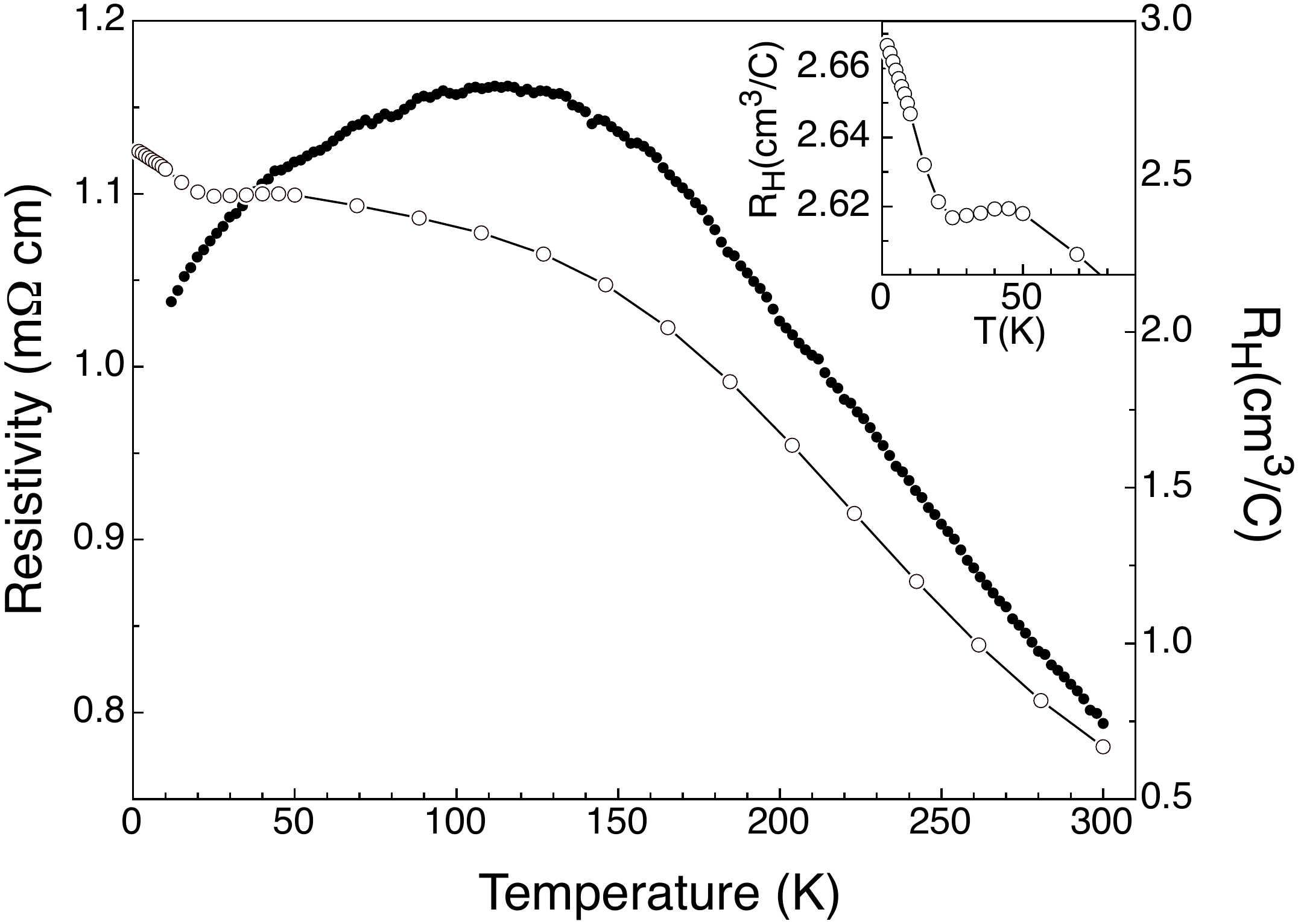}
\caption{The solid points show the resistivity $\rho (T)$ of GdBiPt at zero
magnetic field for a temperature range of 10~K to 300~K. The open circles show
the temperature evolution of the Hall coefficient from 1.8~K to 300K,
revealing a kink well above the 9~K N\'eel temperature (shown in more detail
in the inset).  \label{fig:hall}}
\end{figure}

GdBiPt has a low carrier density ($\sim 3\cdot10^{18}~\mathrm{cm}^{-3}
/ \mathrm{C}$). Figure \ref{fig:hall} shows that there is a gradual increase in the
Hall coefficient as the temperature is reduced, with a clear kink near
25~K. The Hall coefficient was measured using a Quantum Design PPMS, which was also used for the
specific heat measurements.
CeBiPt also shows such a kink followed by a stronger increase of $R_\mathrm{H}$.
In CeBiPt this kink appears at the transition temperature
$T_\mathrm{N}$ and was ascribed to the development of a superzone gap in the
ordered state and consequently a reduction of the number of charge
carries \cite{Jung2001}. In GdBiPt a similar kink seems to be
present, however it occurs
around $25$~K which is above $T_\mathrm{N}\sim 9$~K.

\begin{figure}
\includegraphics[scale=0.3]{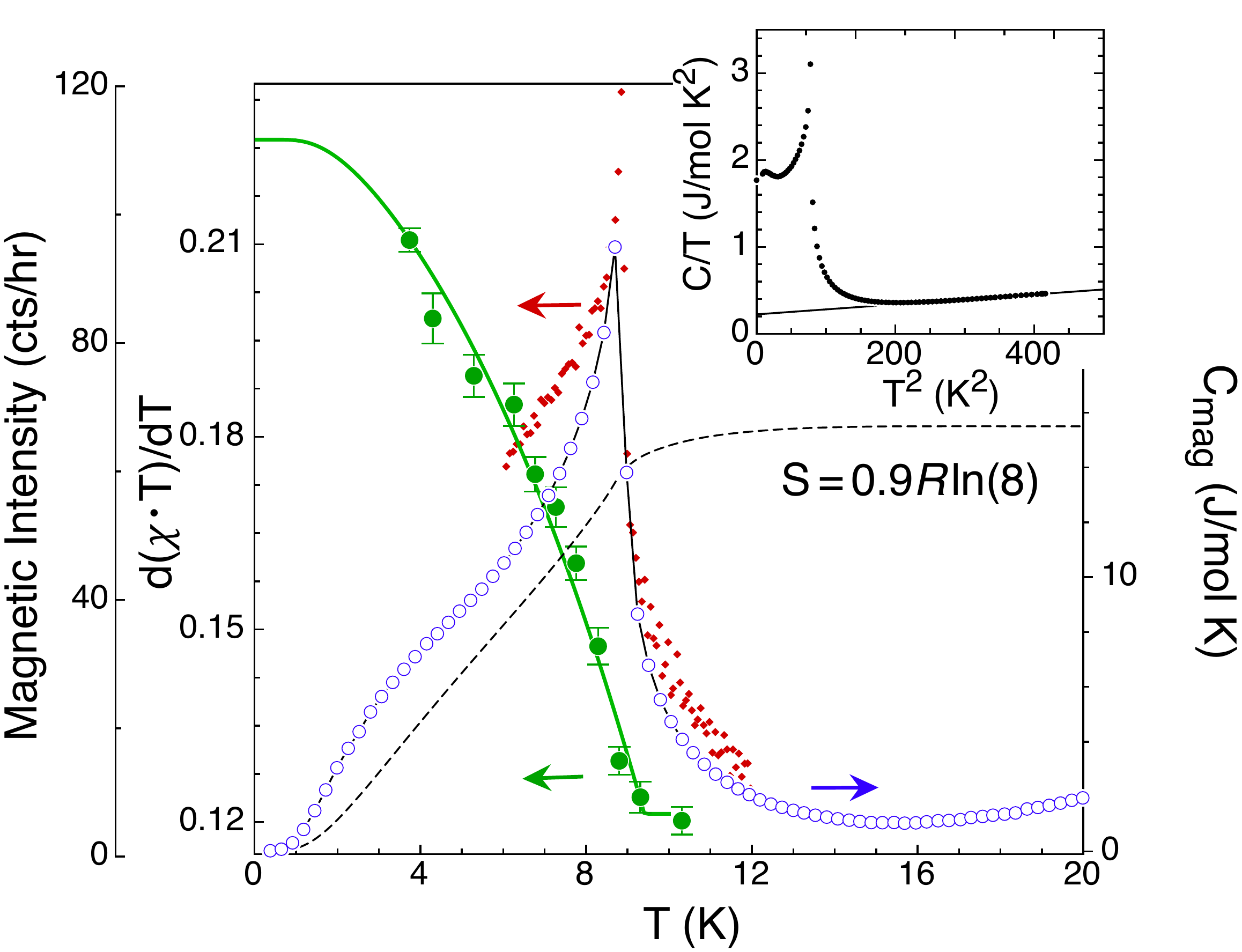}
 \caption{Inset: The specific heat is shown as $C/T$ vs. $T^2$. The solid line
 is a fit to determine the phonon contribution $C_{\mathrm{ph}} = \beta T^3 $
 and the electronic specific heat $C_{el} = \gamma T$. Main figure:  The open
 circles show the magnetic specific heat $C_m =
 C-C_{\mathrm{ph}}-C_{\mathrm{el}}$, solid diamonds show the temperature
 derivative of the magnetic susceptibility $\frac{d}{dT}(\chi T)$. Solid
green circles show the intensity of the first magnetic peak
 $(\frac{1}{2}~\frac{1}{2}~\frac{1}{2} )$ plotted as a function of temperature.
 The solid line is a fit to the square of the magnetic moment, obtained from
numerically solving a Weiss model for a  $J$ of $\frac{7}{2}$.  \label{fig:Cp}}
\end{figure}

For a temperature range of 50 to 300~K, the magnetic susceptibility $\chi$ of
Gd$^{3+}$ shows a Curie-Weiss behaviour with a Curie-Weiss temperature
$\theta_\mathrm{W}$ of $-31.5(3)$~K, and an effective magnetic moment
$\mu_{\mathrm{eff}}$ of $7.97(4) \mu_\mathrm{B}$ consistent with the
$7.94 \mu_\mathrm{B}$
expected  for Gd$^{3+}$. The data were taken in an
applied field of 0.05~T using a Quantum Design VSM squid magnetometer. The
magnetic entropy $S_{\mathrm{mag}}$ shown as the dashed line reaches 
$0.9 R \ln(8)$ at $T_{\mathrm{N}}$ indicative of the absence of frustration in
contrast to the predictions of \cite{Khmelevskyi2012}. Here $S_{\mathrm{mag}}$  was calculated by integrating the
magnetic specific heat  $C-C_{\mathrm{ph}}-C_{\mathrm{el}}$ after
subtracting the phonon $C_{\mathrm{ph}}$ and electronic contributions $C_{\mathrm{el}}$, respectively.
Fig.\ref{fig:Cp} also shows that $\frac{d}{dT}(\chi T)$ exhibits 
a peak at 8.5~K which confirms the antiferromagnetic
ordering with a N\'eel temperature $T_\mathrm{N}$ of  8.5~K. In fact, all three
measurements: Specific heat $C_p(T)$, electrical resistivity
$\frac{d }{d T}\rho(T)$ (not shown), as well as the magnetic susceptibility
$\frac{d}{dT}(\chi T)$, show discontinuities at the same critical temperature
$T_{\mathrm{N}}$, giving evidence to the high quality of our
samples~\cite{Fisher1968a, Fisher1962a}.

At fit to a straight line of  $C/T$ as a function of $T^2$ for
temperatures above 15~K yields a $C_{\mathrm{ph}} = \beta T^3$ with a
\mbox{$\beta$ of $2.9(2)\times 10^{−4}$J /mol K$^4$}. This value of
$\beta$ corresponds to a Debye
temperature $\theta_\mathrm{D}$ of 188(5)~K. 
The same fit results in Sommerfeld coefficient $\gamma$ of only \mbox{2~mJ/mol
  K$^{2}$}, which is low for a metallic compound containing heavy elements such as Gd and Bi. In contrast, the
heavy Fermion YbBiPt shows a $\gamma$ of \mbox{8~J/mol K$^{2}$}, which was
assigned to low lying crystal field levels \cite{Robinson1994}. Since in GdBiPt
the angular momentum $L$ of the $4f^7$ configuration is zero, crystal fields are
not expected to play a significant role. Consequently, we should observe the full magnetic
moment of the Gd$^{3+}$ ion. This is supported by the 0.9~R\,ln(8) entropy
release observed in the phase transition.

\begin{figure}
\includegraphics[width=\linewidth]{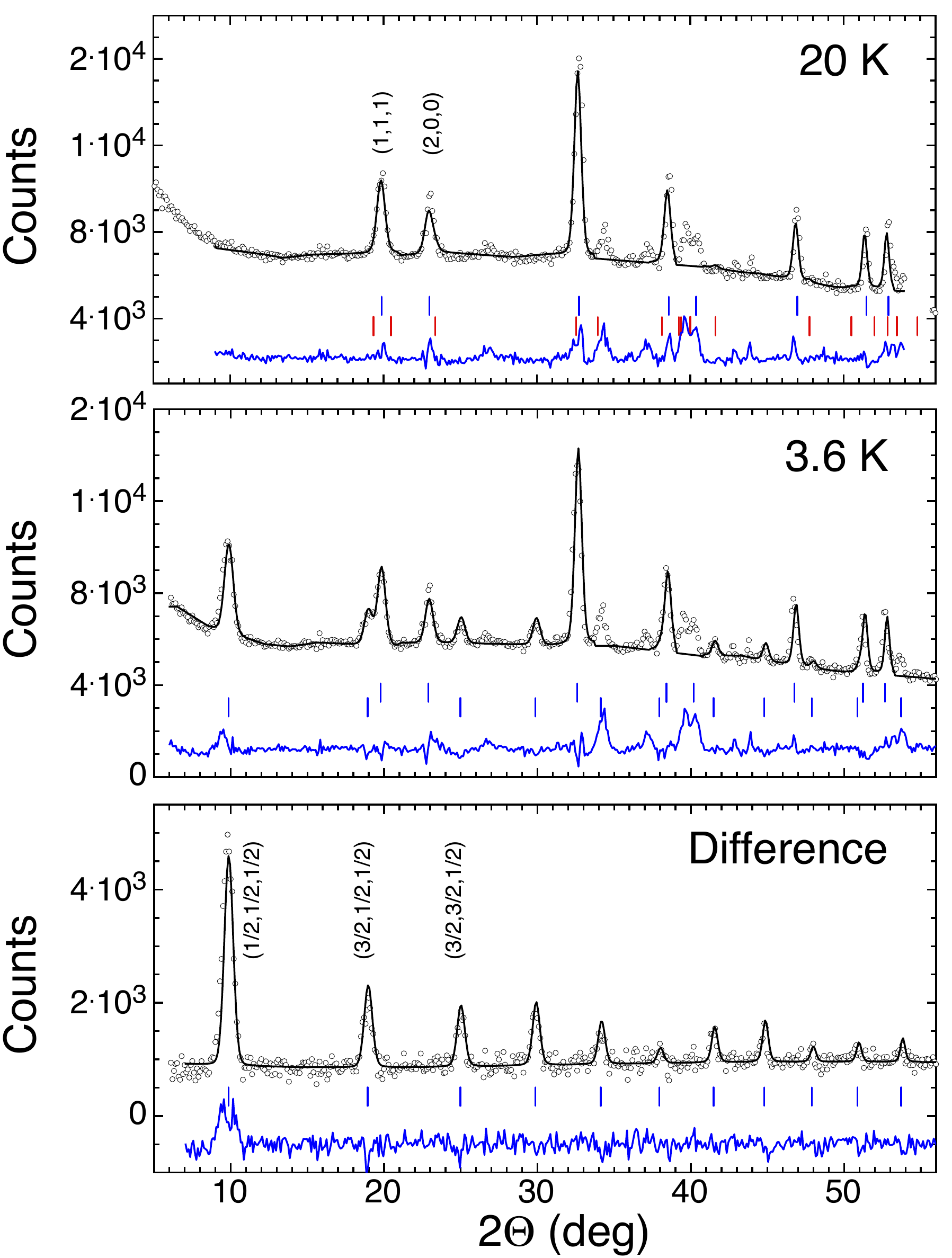}
\caption{Neutron powder diffraction patterns for GdBiPt taken above (20~K, top
panel) and below (3.6~K, middle panel) the N\'eel temperature. The bottom panel
emphasises the form of the magnetic scattering by showing the difference between
the 20~K and 3.6~K patterns. The solid line through the data is a fit (described
in the text) while the solid line below each pattern shows the residuals. In the
20~K pattern (top), the upper set of Bragg markers are for the nuclear
contribution from GdBiPt. The second row indicates the position of Bi
flux. In the
3.6~K pattern (middle), the first row of Bragg markers is the nuclear
contribution, and the bottom row is the magnetic contribution. As the difference
pattern (bottom) only has magnetic peaks, the Bragg markers are for the magnetic
pattern.\label{fig:neutron}} \end{figure}


Our GdBiPt crystals were grown from non-enriched Gd containing the natural
abundance of the different Gd-isotopes which lead to an extreme absorption cross section of GdBiPt~\cite{Ryan2008}.  In order to be still able to
carry out our neutron diffraction experiment,
we used a thinly dispersed sample on a large flat Si sample plate with a
very low background (for details see \cite{SOM,Ryan2008}).
The neutron diffraction pattern in the top panel of Figure~\ref{fig:neutron} was
taken at 20~K, well above the N\'eel temperature. It therefore shows only nuclear
reflections which can be indexed with the MgAsAg-type fcc structure.
On cooling below $T_\mathrm{N}$ to 3.6~K the gadolinium moments order and several magnetic
reflections appear in the middle panel of Figure~\ref{fig:neutron}. All of the
magnetic peaks can be indexed as ($\frac{2n-1}{2} \, \frac{2n-1}{2}
\,\frac{2n-1}{2}$) with $n=$1, 2, ..., indicating that the magnetic
unit cell is doubled along the (1~1~1) direction of the crystallographic unit cell.

\begin{table}[H]
\label{tab:ireps}
\caption{Real (BASR) and imaginary (BASI) components
  of the basis vectors for the two permitted commensurable 
magnetic structures obtained from BasIreps for the space group
$F\overline{4}3m$, an ordering wave vector $\mathbf{k}$ of
$[\frac{1}{2},\frac{1}{2},\frac{1}{2}]$, and Gd$^{3+}$ sitting on
the $4c$ crystallographic site. }
\begin{ruledtabular}
\begin{tabular}{c|c|cc}
& Set 1&  \multicolumn{2}{c}{Set 2}\\[1ex] 
\hline
& & &\\
BASR &   (1 1 1)   &  (1 -0.5 -0.5) & (-0.5 1 -0.5)\\[1ex] 
BASI & (0 0 0) &  (0 -0.866 0.866) & (-0.866 0 0.866)\\[1ex] 
\end{tabular}
\end{ruledtabular}
\end{table}

Plotting the intensity of the first magnetic peak against temperature
(Figure~\ref{fig:Cp}) and fitting it  reveals a N\'eel temperature of 9.4(1)~K, slightly higher
than derived earlier from heat capacity and susceptibility. The
$\mathbf{k}$-vector $\mathbf{k}_1=[\frac{1}{2},\frac{1}{2},\frac{1}{2}]$ of this
type-II antiferromagnetic structure  belongs to a star containing three more
elements  $\mathbf{k}_2=[-\frac{1}{2},\frac{1}{2},\frac{1}{2}]$,
$\mathbf{k}_3=[-\frac{1}{2},-\frac{1}{2},\frac{1}{2}]$ and
$\mathbf{k}_4=[\frac{1}{2},-\frac{1}{2},\frac{1}{2}]$,which are equivalent due to
the cubic symmetry. We then used the BasIreps program, which is part of the
Fullprof Suite \cite{Rodriguez-Carvajal1993a}) to
find the basis functions of the irreducible representations of the
$F\overline{4}3m$ space group with $\bm{k}$=[$\frac{1}{2} \, \frac{1}{2} \,
\frac{1}{2}$]. This symmetry allows two sets of basis
functions whose real and imaginary components are listed in Tab.~\ref{tab:irep}.

For the basis functions listed in Tab.~\ref{tab:ireps} the magnetic
moment is given by:
\begin{equation}
\bold{S}=C \cdot [\bold{BasR} +i~ \bold{BasI}]
\label{eq:moment}
\end{equation}
The two basis functions of set 2 represents the two racemic structures
possible. Due to the fact that we used powder these are
indistinguishable in the refinement and we are left with a single
parameter $C$ as refinable quantity.

The first set of basis functions places the gadolinium moments along the body
diagonal of the cubic structure. However the ($\frac{1}{2} \, \frac{1}{2} \,
\frac{1}{2}$) peak is forbidden for this set since three of the four equivalent  
($\frac{1}{2} \, \frac{1}{2} \,\frac{1}{2}$) peaks are systematically 
absent due to the translational symmetry of the space group (face 
centered), and the fourth is absent due to the magnetic polarization 
factor for neutron scattering. However, it is clear from the difference pattern in
figure~\ref{fig:neutron} that this is the strongest of the observed magnetic
peaks. This allows us to rule out the first set of basis functions.

A refinement of the second set of basis functions contains two equivalent basis
vectors, of which the first was chosen for the refinement. The 3.6~K pattern
returns a Gd magnetic moment of $6.6(7) \mu_\mathrm{B}$ which corresponds to a moment of
$7.6(5) \mu_\mathrm{B}$ at 0~K, which is comparable to value of  $7.55
\mu_\mathrm{B}$ reported for single crystal Gd~\cite{Will1964}.
The difference pattern in Figure \ref{fig:neutron} was also refined and gave
the same $6.7(6) \mu_\mathrm{B}$for the Gd moment at 3.6~K.

Previous resonant magnetic X-ray scattering experiments~\cite{Kreyssig2011}, were
unable to determine the direction of the magnetic moment of GdBiPt. Their attempts to
refine the actual moment direction were inconclusive as they had several
sizeable magnetic domains within the $\sim0.5 \, \times \, 0.5$~mm$^2$ beam
footprint that led to incomplete averaging over directions.
By working with a powder and a much larger ($\sim2.5 \, \times \, 8$~cm$^2$) beam
footprint, domain averaging is complete in our data permitting a full analysis
of the peak intenisties and allowing us to determine the magnetic structure.
Complex (e.g.  cycloidal) ordering was deemed to be incompatible
with the XRMS data~\cite{Kreyssig2011}, and since we detected
no other magnetic scattering down to $2 \theta = 4^{\circ}$, ($q \sim
0.33$\AA$^{-1}$), we can directly rule out long-period modulations of the
magnetic structure with periods less than about 19~\AA\ (about three lattice
spacings). Longer-period modulations would yield satellites around the magnetic
peaks which are also absent. We conclude that GdBiPt adopts a simple
collinear type II antiferromagnetic structure.
The magnetic unit cell is eight times larger than the crystallographic unit
cell, as the $\bm{k}$=[$\frac{1}{2} \, \frac{1}{2} \, \frac{1}{2}$] propagation
vector doubles all three crystallographic axes.  The magnetic moments form
ferromagnetic sheets which are stacked antiferromagnetically along the $[111]$
body diagonal (Figure \ref{fig:sheets}). The same propagation vector is found
for the vanadium doped half-Heusler compound CuMnSb~\cite{Halder2011}, but not
for CeBiPt which orders as a type I AFM with a propagation vector of $[100]$
\cite{Wosnitza2006}. The evaluation of the magnetic moment direction with the
program BasIreps suggests a common, single $\mathbf{k}$-vector structure with the
moments perpendicular to the space-diagonal. Our results make
GdBiPt a srong candidate material for an AFTI.

The results presented here suggest a similar structure to that proposed
by Mong {\it et al.}\cite{Mong2010}, with an observed spin arrangement that results in
strong spin-orbit interaction along the space diagonal. This leads to a path
asymmetry for inter ferromagnetic plane hopping between non-magnetic sites.  In
conclusion, given its spin-structure, GdBiPt is therefore a promising candidate for an
antiferromagnetic topological insulator.

The research at McGill and UdeM received support from the Natural Sciences and
Engineering Research Council of Canada (Canada) and Fonds Qu\'eb\'ecois de
la Recherche sur la Nature et les Technologies (Qu\'ebec).
ADB and YM are also supported by the Canada Research Chair Foundation.
The neutron diffraction measurements were made at the Canadian Neutron Beam
Centre, Chalk River, Ontario.
\bibliography{GdBiPt}
\end{document}


\section{Sample growth}
Single crystals where grown using Bi flux. Gd, Bi and Pt of hight purity where placed in a ceramic crucible in the ratio 1:15:1 which was then sealed in a quartz ampoule under argon atmosphere. The melt was kept at  $1200\,^{\circ}\mathrm{C}$ for two days and then cooled down to $550\,^{\circ}\mathrm{C}$  over a week. After two days the ampoules where then taken out of the furnace and centrifuged to separate the flux from the crystals.

\section{Crystal structure of GdBiPt}
Neutron diffraction experiments were carried out on the C2 multi-wire
powder diffractometer ({\mbox DUALSPEC}) at the NRU reactor of the Canadian Neutron Beam Centre in
Chalk River, Ontario. We used a large-area flat-plate scattering geometry
that reduces the effects of the extreme absorption cross-section of natural
gadolinium and permits neutron diffraction on regular samples
(no isotopic separation) at thermal wavelengths\cite{Ryan2008}.
To prepare the flat-plate samples for the neutron diffraction measurements,
$\sim 400$~mg (about a $1/e$ absorption thickness) of finely powdered material
was spread across a 2.4~cm$\times$8~cm area on a 600~$\mu$m thick single-crystal
silicon wafer and immobilised using a 1\% solution of GE-7031 varnish
in toluene/methanol (1:1) \cite{Ryan2008}. A second silicon wafer was
used as a cover. The two plates were mounted in an aluminium frame
and loaded into a a closed-cycle refrigerator
with the sample in a partial pressure of helium to ensure thermal uniformity.
The plate was oriented with its surface normal to the incident
neutron beam to maximize the total flux onto the sample and
the measurements were made in transmission mode. A neutron wavelength
$\lambda$ of 1.3286(1)~\AA\ 
was used as no long-period antiferromagnetic ordering modes were expected.
We carried out full-pattern magnetic and structural refinements using the
FullProf/WinPlotr suite \cite{Rodriguez-Carvajal1993a, Roisnel2001a} with neutron scattering
length coefficients for natural Gd tabulated by Lynn and
Seeger \cite{Lynn1990} As all of the major magnetic reflections occurred below
35$^\circ$,
no absorption correction was applied, however the data were truncated at
$2 \theta \, = \, 54^\circ$ to minimise any possible impact of angle-dependent
absorption effects.

The room-temperature powder X-ray diffraction data used in the powder
refinement was collected on a
PANalytical X'Pert Pro diffractometer with a linear X'Celerator
detector and Cu $K_{\alpha_1}$ radiation in the 2$\Theta$ range from
20$^{\circ}$ to 120$^{\circ}$. The step size was 0.0084$^{\circ}$ and
the total collection time was 12 hours and 54 minutes. For this
experiment, a few GdPtBi single crystals were finely ground and deposited on the Si single crystal zero-background sample holder.

\begin{table}[H]
\caption{\label{tab:table1} Comparison of the $\chi^2$ obtained from
  refining the powder data for the three different possible
  occupation of the crystallographic sites.}
\begin{ruledtabular}
\begin{tabular}{lcccc}
site&&Order reported in~\cite{Kreyssig2011}&Gd on unique site& Bi on unique site\\[1ex]
\hline
$0,0,0$ ~$4a$&unique site  &Pt & Gd&Bi \\[1ex] 
$\frac{1}{4},\frac{1}{4},\frac{1}{4}$~$4c$& & Gd & Bi&Pt\\[1ex] 
$\frac{1}{2},\frac{1}{2},\frac{1}{2}$~$4b$&ordered vacancy& & &\\[1ex] 
$\frac{3}{4},\frac{3}{4},\frac{3}{4}$~$4d$& & Bi & Pt&Gd\\[2ex]
$\chi^2$&Neutrons&23.7&86.1&32.6\\
&X-rays&10.4&26.3&14.9\\
&Combined&25.1&74.8&34.79\\
\end{tabular}
\end{ruledtabular}
\end{table}

\begin{table}[H]
\caption{\label{tab:neutronresults}Summary of the parameters which were obtained
  from refining the powder diffraction spectra collected at 20,
  3.6~K, and the difference pattern.}
\begin{ruledtabular}
\begin{tabular}{lccc}
& 20~K & 3.6~K & diff\\[1ex] 
\hline
& & &\\
a (\AA)& 6.681(4) & 6.67(9) & 6.68(8)\\[1ex] 
$\mu_{\mathrm{Gd}}$ ($\mu_\mathrm{B}$) & 0 & 6.6(7) & 6.7(6)\\[1ex] 
Rp & 2.20& 2.39 & 9.77\\[1ex] 
Rwp & 2.92 & 3.23 & 12.7\\[1ex] 
Rexp & 1.31 & 1.3 & 3.08\\[1ex] 
$\chi^2$ & 4.99 & 6.16& 17\\[1ex] 
\end{tabular}
\end{ruledtabular}
\end{table}

Room-temperature single crystal X-ray diffraction data were collected
on a STOE IPDSII diffractometer with the Mo$K_{\alpha}$ radiation in the whole
reciprocal sphere. A numerical absorption correction was based on the
crystal shape that was originally derived from the optical face
indexing but later optimized against equivalent reflections using the
STOE X-Shape software \cite{STOEandCieGmbH1997}. Structural refinement was performed using the SHELXL program \cite{Sheldrick2008} (Tables III-V).

\begin{figure}[H]
\centering
\includegraphics[width=\linewidth]{GdBiPt_SOM_Fig1.pdf}
\caption{View of the two possible atom arrangements in a racemic
structure which can be transform into each other through the inversion
symmetry. The Pt atoms are shown in orange, the Bi in grey, and the Gd
in blue.
  \label{fig:racemic}}
\end{figure}

\begin{table}[H]\footnotesize
\caption{\label{tab:Xray1}Refinement of the GdBiPt single crystal X-ray data.}
\begin{ruledtabular}
\begin{tabular}{lcc}
Empirical formula & GdPtBi \\[1ex] 
Formula weight  & 561.32\\[1ex]
Temperature & 293(2) K\\[1ex]
Wavelength & 0.71073~\AA\\[1ex]
Space group & $F\overline{4}3m$\\[1ex]
Unit cell dimensions	& $a = 6.6772(8)$~\AA \ \	$\alpha= 90^\circ$\\[1ex]
                                 &$b = 6.6772(8)$~\AA \ \	$\beta= 90^\circ$\\[1ex]
                                 &$c = 6.6772(8)$~\AA \ \	$\gamma = 90^\circ$\\[1ex]
Volume& 297.70(6)~$\AA^3$\\[1ex]
Z & 4\\[1ex]
Density (calculated) & 12.524 Mg/m$^3$\\[1ex]
Absorption coefficient & 127.510 mm$^{-1}$\\[1ex]
F(000) & 900\\[1ex]
Theta range for data collection	& 5.29$^\circ$ to 34.80$^\circ$\\[1ex]
Index ranges & $-10\leq h \leq 10$, $-10 \leq k \leq 10$, $-8\leq l
\leq 10$\\[1ex]
Reflections collected	 & 1285\\[1ex]
Independent reflections &	90 [R(int) = 0.0856]\\[1ex]
Completeness to theta = 34.80$\circ$ &	100.0~\% \\[1ex]
Refinement method &	Full-matrix least-squares on F$^2$\\[1ex]
Data / restraints / parameters &	90 / 0 / 5\\[1ex]
Goodness-of-fit on F$^2 $&	1.132\\[1ex]
Final R indices [I~$>$~2$\sigma(I)$]	& R1 = 0.0241, wR2 = 0.0511\\[1ex]
R indices (all data) &	R1 = 0.0241, wR2 = 0.0511\\[1ex]
Absolute structure parameter &	-0.13(2)\\[1ex]
Extinction coefficient	 & 0.0152(12)\\[1ex]
Largest diff. peak and hole	& 3.058 and -1.094 e.\AA$^3$\\[1ex]
\end{tabular}
\end{ruledtabular}
\end{table}

\begin{table}[H]\footnotesize
\caption{\label{tab:Xray3}Bond lengths in units of [\AA] obtained from
  refining the single crystal X-ray data.}
\begin{ruledtabular}
\begin{tabular}{lcc}
Gd-Pt 	        &2.8913(3)\\[1ex] 
Gd-Pt\#1 	&2.8913(4)\\[1ex] 
Gd-Pt\#2 	&2.8913(4)\\[1ex] 
Gd-Pt\#3 	&2.8913(4)\\[1ex] 
Gd-Bi\#3 	&3.3386(4)\\[1ex] 
Gd-Bi\#4 	&3.3386(4)\\[1ex] 
Gd-Bi\#1 	&3.3386(4)\\[1ex] 
Gd-Bi\#5 	&3.3386(4)\\[1ex] 
Gd-Bi\#2 	&3.3386(4)\\[1ex] 
Gd-Bi\#6 	&3.3386(4)\\[1ex] 
Bi-Pt 	        &2.8913(3)\\[1ex] 
Bi-Pt\#7 	&2.8913(3)\\[1ex] 
Bi-Pt\#8 	&2.8913(3)\\[1ex] 
Bi-Pt\#9 	&2.8913(3)\\[1ex] 
Bi-Gd\#10 	&3.3386(4)\\[1ex] 
Bi-Gd\#8 	&3.3386(4)\\[1ex] 
Bi-Gd\#11 	&3.3386(4)\\[1ex] 
Bi-Gd\#7 	&3.3386(4)\\[1ex] 
Bi-Gd\#12 	&3.3386(4)\\[1ex] 
Bi-Gd\#9 	&3.3386(4)\\[1ex] 
Pt-Bi\#1 	&2.8913(3)\\[1ex] 
Pt-Gd\#9 	&2.8913(3)\\[1ex] 
Pt-Gd\#8 	&2.8913(3)\\[1ex] 
Pt-Gd\#7 	&2.8913(3)\\[1ex] 
Pt-Bi\#3 	&2.8913(3)\\[1ex] 
Pt-Bi\#2 	&2.8913(3)\\[1ex] 
\end{tabular}
\end{ruledtabular}
\end{table}

\begin{table}[H]
\caption{\label{tab:Xray4}Anisotropic displacement parameters  in
  units of [\AA$^2\cdot$~10$^3$] obtained by refining the single crystal
  X-ray data.  The anisotropic displacement factor exponent takes the form:  $-2\mathrm{p}^2[ \mathrm{h}^2 \mathrm{a}\cdot2\mathrm{U}^{11} + ...  + 2\mathrm{hk~a}\cdot \mathrm{b}\cdot \mathrm{U}^{12} ]$}
\begin{ruledtabular}
\begin{tabular}{lccccccc}
&U$^{11}$ & U$^{22}$ & U$^{33}$ & U$^{23}$ & U$^{13}$ & U$^{12}$\\[1ex] 
\hline
& & &\\
Gd       & 0.0102(7)  & 0.0102(7) & 0.0102(7)   & 0 & 0 & 0\\[1ex] 
Bi 	 & 0.0035(4)  & 0.0035(4) & 0.0035(4)   & 0 & 0 & 0\\[1ex] 
Pt 	 & 0.0091(4)  & 0.0091(4) & 0.0091(4)   & 0 & 0  &0\\[1ex] 
\end{tabular}
\end{ruledtabular}
\end{table}

\section{Supplemental figure of the magnetic structure of GdBiPt}

\begin{figure}[H]
\centering
\includegraphics[width=\linewidth]{GdBiPt_SOM_Fig2.pdf}
\caption{View of the magnetic structure of GdBiPt with the $[111]$
body diagonal pointing to the right. The moments are perpendicular to the body diagonal, and form alternating
ferromagnetic sheets. The Gd atoms and moments are shown in blue, 
Bi atoms are grey, and Pt atoms are yellow.
\label{fig:magstructure}}
\end{figure}

\bibliography{GdBiPt.bib}